\input harvmac

\def\frac#1#2{{#1\over#2}}

\def\half{\frac12}

\def\d{\partial}

\def\inbar{\,\vrule height1.5ex width.4pt depth0pt}
\def\IC{\relax\hbox{$\inbar\kern-.3em{\rm C}$}}
\def\IR{\relax{\rm I\kern-.18em R}}
\def\IP{\relax{\rm I\kern-.18em P}}
\def\IZ{\relax{\rm I\kern-.18em Z}}

\def\VV{{\cal V}}
%
%

\def \sinh{{\rm sinh}}
\def \cosh{{\rm cosh}}


\Title{
\rightline{hep-th/0202183}}
{\vbox{\centerline{Comments on cosmological RG flows}}}
\medskip
\centerline{\it Riccardo Argurio}
\bigskip
\centerline{Institute for Theoretical Physics}
\centerline{G\"oteborg University and Chalmers University of Technology}
\centerline{412 96 G\"oteborg, Sweden}
\bigskip
\centerline{\tt argurio@fy.chalmers.se}
\bigskip\bigskip
\noindent
We study cosmological backgrounds from the point of view of the dS/CFT 
correspondence and its renormalization group flow extension.
We focus on the case where gravity is coupled to a single scalar with a
potential. Depending on the latter, the scalar can drive both inflation 
and the accelerated expansion (dS) phase in the far future.
We also comment on quintessence scenarios, and flows familiar from 
the AdS/CFT correspondence. We finally make a tentative embedding
of this discussion in string theory where the scalar is the dilaton
and the potential is generated at the perturbative level.

\vfill

\Date{02/02}

\lref\carroll{S.~M.~Carroll,
``The cosmological constant,''
Living Rev.\ Rel.\  {\bf 4}, 1 (2001)
[arXiv:astro-ph/0004075]; 
``TASI lectures: Cosmology for string theorists,''
arXiv:hep-th/0011110.}
\lref\banks{T.~Banks,
``Cosmological breaking of supersymmetry or little Lambda goes back to  
the future. II,''
arXiv:hep-th/0007146.
}
\lref\witten{E.~Witten,
``Quantum gravity in de Sitter space,''
arXiv:hep-th/0106109.
}
\lref\stromdscft{A.~Strominger,
``The dS/CFT correspondence,''
JHEP {\bf 0110}, 034 (2001)
[arXiv:hep-th/0106113];
M.~Spradlin, A.~Strominger and A.~Volovich,
``Les Houches lectures on de Sitter space,''
arXiv:hep-th/0110007.
}
\lref\stromrg{A.~Strominger,
``Inflation and the dS/CFT correspondence,''
JHEP {\bf 0111}, 049 (2001)
[arXiv:hep-th/0110087];
V.~Balasubramanian, J.~de Boer and D.~Minic,
``Mass, entropy and holography in asymptotically de Sitter spaces,''
arXiv:hep-th/0110108.
}
\lref\adscft{For a review, see
O.~Aharony, S.~S.~Gubser, J.~Maldacena, H.~Ooguri and Y.~Oz,
``Large N field theories, string theory and gravity,''
Phys.\ Rept.\  {\bf 323}, 183 (2000)
[arXiv:hep-th/9905111].
}
\lref\skto{K.~Skenderis and P.~K.~Townsend,
``Gravitational stability and renormalization-group flow,''
Phys.\ Lett.\ B {\bf 468}, 46 (1999)
[arXiv:hep-th/9909070].
}
\lref\karch{O.~DeWolfe, D.~Z.~Freedman, S.~S.~Gubser and A.~Karch,
``Modeling the fifth dimension with scalars and gravity,''
Phys.\ Rev.\ D {\bf 62}, 046008 (2000)
[arXiv:hep-th/9909134].
}
\lref\gppz{L.~Girardello, M.~Petrini, M.~Porrati and A.~Zaffaroni,
``Novel local CFT and exact results on perturbations of N = 4 super  
Yang-Mills from AdS dynamics,''
JHEP {\bf 9812}, 022 (1998)
[arXiv:hep-th/9810126].
}
\lref\freed{D.~Z.~Freedman, S.~S.~Gubser, K.~Pilch and N.~P.~Warner,
``Renormalization group flows from holography supersymmetry and a  
c-theorem,''
Adv.\ Theor.\ Math.\ Phys.\  {\bf 3}, 363 (1999)
[arXiv:hep-th/9904017].
}
\lref\behrndt{K.~Behrndt,
``A non-singular infrared flow from D = 5 gauged supergravity,''
Phys.\ Lett.\ B {\bf 487}, 30 (2000)
[arXiv:hep-th/0005185].
}
\lref\gppzb{L.~Girardello, M.~Petrini, M.~Porrati and A.~Zaffaroni,
``The supergravity dual of N = 1 super Yang-Mills theory,''
Nucl.\ Phys.\ B {\bf 569}, 451 (2000)
[arXiv:hep-th/9909047].
}
\lref\quint{S.~Hellerman, N.~Kaloper and L.~Susskind,
``String theory and quintessence,''
JHEP {\bf 0106}, 003 (2001)
[arXiv:hep-th/0104180];
W.~Fischler, A.~Kashani-Poor, R.~McNees and S.~Paban,
``The acceleration of the universe, a challenge for string theory,''
JHEP {\bf 0107}, 003 (2001)
[arXiv:hep-th/0104181].
}
\lref\tsva{A.~A.~Tseytlin and C.~Vafa,
``Elements of string cosmology,''
Nucl.\ Phys.\ B {\bf 372}, 443 (1992)
[arXiv:hep-th/9109048].
}
\lref\silv{E.~Silverstein,
``(A)dS backgrounds from asymmetric orientifolds,'' 
arXiv: hep-th/0106209.
}
\lref\larsen{F.~Larsen, J.~P.~van der Schaar and R.~G.~Leigh,
``de Sitter holography and the cosmic microwave background,''
arXiv:hep-th/0202127.
}
\lref\hull{C.~M.~Hull,
``Timelike T-duality, 
de Sitter space, large N gauge theories and  topological field theory,''
JHEP {\bf 9807}, 021 (1998)
[arXiv:hep-th/9806146].
}
\lref\hullb{C.~M.~Hull,
``de Sitter space in supergravity and M theory,''
JHEP {\bf 0111}, 012 (2001)
[arXiv:hep-th/0109213]. 
}
\lref\townq{P.~K.~Townsend,
``Quintessence from M-theory,''
JHEP {\bf 0111}, 042 (2001)
[arXiv:hep-th/0110072].
}
\lref\huto{C.~M.~Hull,
``Domain wall and de Sitter solutions of gauged supergravity,''
JHEP {\bf 0111}, 061 (2001)
[arXiv:hep-th/0110048].
}
\lref\bala{V.~Balasubramanian, P.~Horava and D.~Minic,
``Deconstructing de Sitter,''
JHEP {\bf 0105}, 043 (2001)
[arXiv:hep-th/0103171].
}
\lref\kallosh{R.~Kallosh,
``N = 2 supersymmetry and de Sitter space,''
arXiv:hep-th/0109168.
}
\lref\bandin{T.~Banks and M.~Dine,
``Dark energy in perturbative string cosmology,''
JHEP {\bf 0110}, 012 (2001)
[arXiv:hep-th/0106276].
}
\lref\lmm{F.~Leblond, D.~Marolf and R.~C.~Myers,
``Tall tales from de Sitter space. I: Renormalization group flows,''
arXiv:hep-th/0202094.
}
\lref\num{V.~L.~Campos, G.~Ferretti, H.~Larsson, D.~Martelli and B.~E.~Nilsson,
``A study of holographic renormalization group flows in d = 6 and d = 3,''
JHEP {\bf 0006}, 023 (2000)
[arXiv:hep-th/0003151].
}
\lref\boonstra{H.~J.~Boonstra, K.~Skenderis and P.~K.~Townsend,
``The domain wall/QFT correspondence,''
JHEP {\bf 9901}, 003 (1999)
[arXiv:hep-th/9807137].
}
\lref\berg{M.~Berg and H.~Samtleben,
``An exact holographic RG flow between 2d conformal fixed points,''
arXiv:hep-th/0112154.
}

\newsec{Introduction}
Recent experimental data can be interpreted in terms of
the presence in our universe of a positive cosmological 
constant (see \carroll\ for a review). 
Its magnitude is however very small with respect
to natural quantum gravity units, $\Lambda \sim 10^{-120} m_p^4$.
Nevertheless, a positive constant energy density is bound to 
dominate the evolution of the universe in the far future.
The future evolution will thus be very close to a de Sitter universe.
Note that since only the expanding de Sitter universe is relevant
to these considerations, we can assume that it has open spatial
sections. We can thus write the dS metric in the following most
useful form:
\eqn\dsmetric{ds^2=-dt^2+ e^{2Ht}dx_i^2 .}
Here $t$ is the cosmological comoving time and $H$ is the Hubble constant
(which, for de Sitter, is {\it really} constant) associated to $\Lambda$
as $H^2 \sim \Lambda / m_p^2 $. A universe with a positive cosmological
constant will thus have a metric asymptotic to \dsmetric\ for
$t \to \infty$.

An accelerated (de Sitter like) expansion is also present in
the inflationary scenario. This expansion stage is supposed to have
taken place in the very early universe and is useful to explain
various features of the present universe (most notably its flatness).
We can thus suppose that during this stage the universe
also had a metric similar to \dsmetric, however with a different
Hubble constant $H$. For our following considerations, let us just
assume that in contrast with the ``future'' $H$, the ``inflationary'' $H$
is (within a few orders of magnitude) of the Planck scale.
Note here that the inflationary de Sitter phase in the past has not
to be confused with a de Sitter universe near its past infinity:
such a contracting universe is the time reversal of \dsmetric\ and
is not relevant to (traditional) cosmological considerations.
So to cut the story short, the universe has expanding de Sitter phases
both in the past and the future, however there is an enormous 
hierarchy between the two cosmological constants driving
the expansion.

Given these observations based on experimental data, the question
arises of how to incorporate them into string theory. Setting aside
conceptual problems related to quantum gravity in de Sitter space 
\refs{\banks, \witten}, an interesting point of view is drawn from
the analogy with the AdS/CFT correspondence \adscft, where 
it is postulated that a gravitational theory in the bulk of de Sitter spacetime
is dual to a Euclidean CFT situated on its future boundary \stromdscft\
(we consider
here the set up of the correspondence where the metric is
\dsmetric, which covers half of de Sitter
spacetime and reaches only to one boundary). 
A step further in the analogy is to associate the cosmological
time coordinate $t$ with renormalization group flow in the dual CFT 
\stromrg: time evolution in de Sitter corresponds to
scale transformations in the dual CFT.
It is then natural to generalize that beyond pure de Sitter and consider
renormalization group flows in the boundary theory as cosmological
evolution in the bulk. In particular, evolution between two de Sitter
phases is interpreted as the RG flow of the dual theory between 
two fixed points \stromrg, with the UV one (corresponding to the future)
having a much larger central charge than the IR one.

Once this correspondence between cosmological evolution in the bulk
and RG flow in the boundary field theory is stated, it is interesting
to investigate it in many respects. The many AdS/CFT
techniques and concepts developed
in the context of the domain wall/RG flow  correspondence can be
translated into dS language, and differences analyzed. Some of the flows
considered in AdS space, when mapped to dS, can be actually given a
cosmological interpretation. Conversely, cosmological models can be 
analyzed with field theory insight.
The major drawback of dS considerations versus AdS ones, is that in the
latter extended supersymmetry allows us to consider very specific
potentials which entail precise predictions on the field theory side.
In the dS case, for which supersymmetry is necessarily absent (presumably
broken), we are left with generic considerations and toy potentials.

In the following, we will firstly formulate the problem of cosmological
evolution coupled to a scalar with a potential, and cast the equations
of motion in a first order form. This implies that the potential derives
from a prepotential. We then analyze generically the field theory
interpretation of the extrema of the potential and prepotential,
together with some considerations about the scalar field solutions
around the extrema and their interpretation in the dual field theory.
We then consider a few specific flows, both exact ones and others
from the AdS/CFT literature, and comment on their cosmological
implications.
A particular class of flows is the one generated by potentials which
are a sum of exponentials, since they could come from loop corrections
to the string effective action. Their cosmological and RG flow
interpretation is given.
We conclude with a brief discussion.

\newsec{Cosmology with a scalar potential}
We are going to consider the simplest model of cosmological evolution,
gravity coupled to a scalar with potential. 
Although a set up with several scalars is more general, the 
exact analytic flows we want to consider
are more likely to arise when only one scalar is active, as for example in
the AdS flows presented in \refs{\gppzb,\berg}. Moreover we will apply
in section 5
this formalism to the case where the scalar is the dilaton, while the
other moduli are (hopefully) frozen.

We take the spacetime
to be $D+1$ dimensional, and we single out the time coordinate $t$.
The action is:
\eqn\action{S=m_p^{D-1}
\int dt d^Dx \sqrt{-g} \left(R-\half \d \phi^2 - V(\phi)\right).}
The metric is taken to be isotropic, homogeneous and flat:
\eqn\metric{ds^2=-e^{2b(t)}dt^2+ e^{2a(t)}dx_idx_i, \qquad i=1\dots D.}
We include for the moment a lapse function, to be fixed later to 1.
The Einstein equations are thus:
\eqn\einsti{{D(D-1)\over 2}\dot{a}^2 = {1\over 4}\dot{\phi}^2
+\half e^{2b} V \equiv \half e^{2b}\rho,}
\eqn\einstii{-(D-1)\left( \ddot{a}- \dot{a}\dot{b}+{D\over 2}\dot{a}^2
\right)= {1\over 4}\dot{\phi}^2 - \half e^{2b} V \equiv \half e^{2b} p,}
where the energy density is defined by $\rho= -{T^t}_t$ and the pressure
by $p={T^1}_1=\dots={T^D}_D$.
Note that the positivity of the kinetic term assures the null energy
condition $p+\rho\geq 0$, however the dominant energy condition
$p\leq \rho$ is satisfied if and only if the potential is non negative
$V\geq 0$. This is what we will assume here, indeed it will turn
out to be a crucial condition for most of the flows. Note also that
it is the opposite condition that is satisfied for AdS flows.

The equation of motion of the scalar field is:
\eqn\scalareom{\ddot{\phi}+D\dot{a}\dot{\phi}-\dot{b}\dot{\phi} +
e^{2b}\d_\phi V=0.}
This equation together with \einsti\ imply \einstii, so the latter
is redundant.

The action can be rewritten using the reduced variables $a(t)$, $b(t)$ and
$\phi(t)$:
\eqn\actionred{S\propto \int dt e^{Da-b}\left\{ -D(D-1)\dot{a}^2
+\half \dot{\phi}^2 - e^{2b} V \right\}.}
The energy functional vanishes on-shell due the $G_{tt}$ equation, \einsti.

We now wish to rewrite the equations of motion in terms of first
order equations, provided the potential $V$ is of some specified form.
That is automatically provided in a supersymmetric theory, here however
we have to make a specific ansatz. Mimicking the technique of 
\refs{\skto, \karch}, we suppose the potential is the difference
of two squared terms. It turns out that the correct expression
in this case is (see also \lmm\ for a recent similar discussion in de Sitter):
\eqn\prepot{V={D\over D-1} W^2-2 (\d_\phi W)^2.}
The equations \einsti\ and \scalareom\ then are satisfied when:
\eqn\bpsphi{\half \dot{\phi}  = - e^b \d_\phi W, }
\eqn\bpsa{(D-1) \dot{a} =  e^b W.}
We have fixed here a sign ambiguity for $W$. In the following we will
mainly use the $b=0$ gauge.
Let us stress that \prepot\ is motivated only for finding first order
equations from which the equations of motion derive, and not by
positive energy considerations. The energy is indeed always vanishing
on shell.

It is interesting to note that the expression for the potential
\prepot\ is ``upside down'' with respect to its canonical form
in supersymmetric (or, equivalently, AdS) theories. Most notably, 
solutions for which $\phi$ is constant will have a positive
energy density since in that case $V\propto W^2$. This is of course
consistent with the fact that we are interested in spacetimes
which asymptote to de Sitter.

Let us comment briefly on the c-function and the c-theorem. In 
analogy with \refs{\gppz, \freed}, a
c-function has been proposed in \stromrg\
for the flat homogeneous isotropic case which is $c\sim H^{-(D-1)}$,
where $H\equiv \dot{a}$ in our notations
(see also \lmm\ for a recent refinement of the proposal). 
Note that a small cosmological
constant, and hence a small Hubble constant $H$, leads to a large
central charge. Conversely, a cosmological constant of the Planck
scale leads to a central charge of order unity.
The c-theorem follows directly from the Friedmann equations for such 
a cosmology, $\dot{H}\sim -(\rho + p)\leq 0$ in all generality,
thus implying $\dot{c}\geq 0$. This identifies the UV with the future
(which is near the dS boundary) and the IR with the past (the deep
dS ``interior'').

Note that if we enforce $V\geq 0$ and thus $p\leq \rho$, we also have an
upper bound on the variation of $c$. This is however somewhat loose.
Note first that: 
\eqn\cbound{\dot{c} = \half {c \over H} (\rho + p) \leq {c \over H} \rho 
= D(D-1) c H,} 
where we have used \einsti.
If $c$ saturates the bound on its variation, and we write
the ``volume'' of a spacelike section as $v=e^{Da}$, we get $c\sim v^{D-1}$. 
Thus $c$ is allowed to grow even faster than the volume of the spacelike
section of the universe (for $D>2$; for $D=2$, that is 3 dimensional
cosmology, $c$ can at most grow as the volume). In some sense, 
the growth of degrees of freedom in the boundary theory can be 
bigger than the expansion of the volume in Planck units of a spatial section
of the universe.

\newsec{General considerations on the potential}
In this section we make general considerations on the extrema of the
potential, in relation with the flows and cosmologies to which they
correspond.

As it is clear from \bpsa, the Hubble constant is proportional
to the prepotential. If we want our flow (or our cosmology) to go 
between two fixed values of $\phi$, say $\phi_\pm \equiv \phi(t\to
\pm \infty)$, 
we then immediately see that $W(\phi_\pm)\equiv W_\pm$ must be 
extrema, and such that $W_-> W_+> 0 $ to comply with $\dot{H}\leq 0$
and $H>0$ for an expanding universe.
Clearly, the generic situation is when $\phi$ starts rolling down 
from a maximum in $W_-$ and lands to a minimum in $W_+$.
Of course less generic situations can be thought of, where higher
derivatives of $W$ vanish in $\phi_-$ and/or $\phi_+$.

Once stated the above, we have to make sure that the evolution
still seems reasonable in terms of the potential $V$, for instance
we want that the universe settles into a minimum of $V$ in order
to avoid excessive fine tuning.
To do this, we first note that $V$ has two kinds of extrema, those
for which $W'=0$, that is the extrema of the prepotential, and those
for which ${D\over D-1} W= 2 W''$ (in the following we denote by
primes derivatives with respect to $\phi$). Following the argument
of the previous paragraph, our first order equations are best suited 
to handle extrema of the first kind. We would thus like to enforce
as a first condition
that there is no extremum of the second kind in between the two
extrema we are considering.

Secondly, let us consider the second derivative of $V$ evaluated at
an extremum of $W$:
\eqn\maxv{ V''|_{W'=0}=2W''\left({D\over D-1}W-2 W''\right).}
Assuming the flow begins at a (positive) maximum of $W$, then $\phi_-$ is
clearly also a maximum of $V$. The minimum of $W$ is also a 
minimum of $V$ provided ${D\over D-1} W_+> 2 W_+''$. This is obviously
the same as requiring that there is no extremum of the second kind
between $\phi_-$ and $\phi_+$.

Thus the condition for the flow to be acceptable in our framework is that 
${D\over D-1} W> 2 W''$ throughout the flow\foot{Given a potential $V$
one can always try to solve \prepot\ for $W$, with boundary conditions
such that the flow is between two $W'=0$ extrema (see e.g.~\num\ for a 
discussion of this issue in AdS/CFT). However the generic
solution is not analytic and thus beyond the scope of these considerations.}
(another condition is of
course that the potential $V$ is non negative along the flow).

Going now back to the RG interpretation we see that the cosmologically
trivial remark that a scalar field has to roll down the potential and
stabilize eventually at a minimum, translates into the c-theorem
where one starts from a fixed point in the IR (the maximum of both $V$ and
$W$) and then integrates in degrees of freedom at higher and
higher energies to eventually reach a fixed
point in the UV (the minimum of $V$ and $W$). We have here of course
a flow between two fixed points because we suppose that both extrema are
positive. 

The c-theorem in dS language can be written as $H_{IR}>H_{UV}$, or
$\Lambda_{IR}>\Lambda_{UV}$. This is the same as for AdS flows, if one
replaces the cosmological constants with their absolute values.
In AdS however, the potential $V$ is really upside down, and thus
non positive, so that the IR extremum is lower than the UV one.

We can also analyze the behaviour of the scalar field near the UV
extremum. We note immediately that the first order equation \bpsphi\ 
actually picks up one of the two asymptotic solutions of the
scalar equation \scalareom.

Generally, if $V''_+=m^2$ and $V_+=D(D-1)H$, 
the scalar equation \scalareom\
becomes $\ddot{\phi}+D H \dot{\phi}+ m^2 \phi=0$ and it has asymptotic
solutions:
\eqn\asympt{\phi \sim e^{\lambda_\pm t}, \qquad \qquad
\lambda_\pm = -{DH\over 2} \pm\sqrt{ \left({DH\over 2}\right)^2 -m^2 }.}
It is interesting to note that since $\phi_+$ is also an extremum of the
prepotential, \maxv\ implies that:
\eqn\bfbound{m^2=2W_+''(DH-2W_+'')\leq \left({DH\over 2}\right)^2,}
and thus the two roots $\lambda_\pm$ are always real:
\eqn\roots{\lambda_\pm=-{DH\over 2}\pm \left|{DH\over 2}-2W_+''
\right|.}
Note we always have $\lambda_+>-{DH\over 2}>\lambda_-$, and we
can accordingly associate dimensions in the dual CFT as
$\lambda_-=-\Delta H$ and $\lambda_+=-H(D-\Delta)$ (restricting here
to the case $\Delta \geq D/2$).
The situation where the roots are real%
\foot{Actually, having real $\phi$ solutions is equivalent
to be able to reformulate the problem with first order equations and a 
real prepotential.} 
is surprising against
flat space intuition, but saves us from having to consider
complex dimensions and thus an explicitly non unitary boundary CFT \stromdscft.

The conventional wisdom in AdS/CFT is that if the solution is normalizable
in AdS it corresponds to a flow generated by giving a VEV to an operator
of dimension $\Delta$, while if the solution is non normalizable in AdS,
the flow corresponds to deforming the theory by the same operator.
In dS the situation is more tricky, since normalizability is 
defined on a spacelike section. 
However in order to draw an analogy with the AdS case, it seems natural
to keep distinguishing solutions according to their normalizability
along the time direction (this is equivalent to considering
the finiteness of their asymptotic energy, see \larsen\ for a recent related 
discussion in the IR). We thus wish to associate solutions
with $\phi \sim e^{\lambda_+ t}$ with a deformation of the boundary theory
with an operator of dimension $\Delta$, while we associate the solutions
$\phi \sim e^{\lambda_- t}$ with giving a VEV to the same operator.

It is then straightforward to continue the analogy by mapping minima
with $m^2>0$ to relevant operators ($\Delta<D$), extrema with $m^2=
V''_+=0$ to marginal ones, and maxima with $m^2<0$ (actually tachyons)
to irrelevant operators.
Thus the situation considered before where we evolve to a minimum
in the UV is obviously associated with relevant operators in the UV
theory.

The first order equation \bpsphi\ picks up the solution:
\eqn\phifirst{\phi\sim e^{-2W_+'' t}.}
It will be non normalizable in the sense discussed above if $W_+''<
{DH\over 4}$. In this case it corresponds to a (relevant)
deformation of the dual
CFT. If on the other hand ${DH\over 4}<W_+''<{DH\over 2}$, it will
be related to the presence of a VEV.

The marginal case occurs when $V''_+=0$. This can happen for 2 reasons,
either $W_+''=0$ or $2W_+''=DH$. In the first case, the solution
\phifirst\ is actually power behaved, and corresponds to the
deformation by a (presumably not exactly) marginal operator, while
the second case corresponds to the VEV of the marginal operator.

Let us make a last general comment. It is interesting to consider
the evolution of the equation of state $p=w\rho$ of the scalar field.
Near the extrema of the flow, the stress energy tensor is dominated
by the potential, and thus the equation of state approaches $p=-\rho$
as for a cosmological constant. One can ask which is the maximum value
of $w$ along the flow, knowing that it is bounded by 1 since the
potential is non negative.
Inserting \prepot\ and \bpsphi\ into the definitions of $p$ and $\rho$,
one can write:
\eqn\wevol{w={p\over \rho}=-1 + 4{D-1\over D} {{W'}^2\over W^2}.} 
While one can certainly build ad hoc (pre)potentials such that $w$
rises close to 1 at a certain point in the flow, it is striking that
limiting oneself to a simple minimum of $W$ one cannot get over $w=0$.
Indeed, if $\phi$ is sufficiently close to the minimum, we can 
write $W=W_+ + \half W''_+ \phi^2$. The maximum of $w$, which
is attained when $WW''=W'^2$, implies that $\half W''_+ \phi_{max}^2=W_+$.
This leads to ${{W'}^2\over W^2}|_{max}=
{W''\over W}|_{max}
={W''_+\over 2W_+}< {D\over 4(D-1)}$ (by requiring $\phi_+$ to be a 
minimum of $V$), which in turn gives $w<0$.
This is actually an interesting feature, since while in this
model the scalar field is supposed to drive both inflation
and the future accelerating phase, it would be unrealistic if it
was to simulate some kind of matter/radiation at intermediate stages.
On the other hand, it can be shown that if $W$ has a zero of higher
(even) order, the parameters can be tuned so that $w$ rises above 0
at some point of the flow near the minimum. These are flows associated
with deformations of the boundary theory by marginal operators.

\newsec{Some cosmological RG flows}
Due to the absence of supersymmetry, and thus of a theory providing
us with a class of (pre)potentials, we are actually left free to analyze
general forms of potentials and the cosmological flows that they give
rise. 
One could actually even take the opposite approach, to choose an
interesting (analytic) flow and then derive the potential that generates it.

For instance, a most obvious cosmological evolution interpolating
between Hubble constants $H_-$ in the far past and $H_+$ in the
far future is the following:
\eqn\thflow{\dot{a}={H_-+H_+ \over 2}-{H_--H_+ \over 2}\tanh {t\over T},}
where of course $H_->H_+$, and $T$ is the time scale of the evolution.
From \bpsa\ we thus have an expression of $W(t)\equiv W(\phi(t))$.
In combination with \bpsphi\ we can straightforwardly find an equation
for the temporal evolution of $\phi$, $\dot{\phi}=\sqrt{2|\dot{W}|}$,
which is solved by:
\eqn\phievol{\cos {\xi \phi} = - \tanh {t\over T}, \qquad
\qquad \xi^{-2}=T(D-1)(H_--H_+).}
The scalar field evolves from $\phi_-=0$ to $\phi_+=\pi/\xi$.
Plugging \phievol\ into \thflow, we finally get the prepotential:
\eqn\wcos{W(\phi)=(D-1)\left({H_-+H_+ \over 2}+{H_--H_+ \over 2}
\cos {\xi \phi}\right),}
which has a maximum in $\phi_-$ and a minimum in $\phi_+$.
Now in order to satisfy the condition which ensures that we land in a 
minimum of the potential $V$ as discussed after \maxv, we need to
take $T>(DH_+)^{-1}$. 
This condition states that the age of the universe (more
precisely, the characteristic time of the evolution from one
de Sitter phase to another) in such a flow is bigger than the scale
fixed by the late time cosmological constant $\Lambda_+\sim H_+^2$.
That would give a time scale of (roughly) an order of magnitude or so larger
than the assumed age of the universe.
Note however that, in agreement with the general argument after \wevol,
we find that $w<0$ along all the flow (and $w$ comes very close to 0
at a point in the flow if $H_+\ll H_-$). Thus if we suppose that some
form of matter and/or radiation dominates the evolution at some
intermediate stage, without disrupting the general behaviour of the
flow, the timescale $T$ could well be suitably reduced to an acceptable
magnitude.

In the M-theory context a prepotential like \wcos\ has been considered 
for instance in \behrndt, and it can possibly arise in non supersymmetric
set ups. However for a prepotential like \wcos\ to be
cosmologically interesting, one needs to have $H_+\ll H_-$ and thus
to fine tune its minimum to be almost on the $\phi$ axis.

A very similar story applies to cubic prepotentials, where again
one has to fine tune the minimum to be close to the axis. The similarity
derives from the fact that for both potentials the two extrema 
are simple zeroes of $W'$. Also in this case it is straightforward
to find an analytic flow. Thus the simple flow \thflow\
seems to be actually a good representative of a rather general class of flows.

A different class of prepotentials leading to an analytic flow derives
from the analogy with the flow analyzed in \gppzb. In its original
AdS version, this flow represents a mass deformation of $N=4$ SYM away
from its UV fixed point towards pure $N=1$ SYM in the IR. The confining
behaviour of the latter is translated into the singularity at finite 
radius of the (analytic) domain wall solution.

Reverting to our de Sitter framework, we can write this class of potentials
as:
\eqn\wcosh{W=g+h\ \cosh \alpha \phi, }
where $h>0$ and $g+h>0$. The condition for the potential $V$ to be positive
everywhere (and thus to have only one minimum at $\phi=0$) is that
${D-1\over D}2\alpha^2$ must be smaller than 1 if $g\geq 0$ or smaller
than $1+{g\over h}$ if $g<0$.

Choosing the gauge $b=0$, the equations \bpsphi-\bpsa\ become:
\eqn\phicosh{\half \dot{\phi}=-\alpha h\ \sinh \alpha \phi,}
\eqn\acosh{(D-1) \dot{a}=g+h\ \cosh \alpha \phi.}
The solution to \phicosh\ is:
\eqn\solcosh{e^{-\alpha \phi}=\tanh \alpha^2 h t.}
We thus see that time starts at $t=0$ where $\phi$ is infinite and
then continues to $t\to \infty$ where $\phi$ approaches 0.
{}For the metric we get:
\eqn\metsol{e^{a}=e^{{g\over D-1}t} (\sinh 2\alpha^2 h t)^{1 \over 2\alpha^2
(D-1)}.}
Thus for $t\to \infty$ we recover the behavior $a \simeq {g + h\over D-1}t
\equiv H_+ t$, that is the late time de Sitter expansion.
However for $t\to 0$ in the past, the metric becomes singular
as:
\eqn\quibe{e^{a} \sim t^\gamma, \qquad\qquad \gamma={1 \over 2\alpha^2(D-1)}.}
This is a Big Bang dominated by matter with equation of state
given by $w=-1+ {2\over D\gamma}=-1+4\alpha^2{D-1\over D}$.
We thus have $-1<w\leq 1$, but the precise value depends on the model.

The dS version of the flow of \gppzb\ thus represents a cosmology
which starts with a Big Bang driven by a specific though model
dependent form of matter, and then evolves towards a de Sitter phase.
A dual theory which confines in the IR therefore seems to be 
unsuitable to describe primordial inflation.

A possibility which is in a certain sense opposite to the above one
is when we have a potential with an exponential runaway behavior
towards zero. This is a well-studied class of potentials in the context of
quintessence cosmological models (see \quint\ for a string theory
point of view on quintessence). 
In our framework, the simplest such potential derives from a prepotential
which consists of a single exponential. There are then no extrema and
in the RG flow interpretation there are no fixed points.
If however we take two exponential terms, $W$ can have a maximum and the
RG flow has then a fixed point in the IR but no fixed point in the UV.

A generic exponential prepotential which has a maximum and a runaway
behavior is:
\eqn\potexp{W=\mu e^{\alpha \phi}- \nu e^{\beta \phi},\qquad \qquad \alpha
< \beta,}
where all the constants are positive. Depending on the parameters, 
the potential $V$ derived from \prepot\ has one of the following three
behaviors: i) a positive maximum followed by a negative minimum and then 
blows up to $+\infty$, ii) a positive maximum and then blows up
to $-\infty$ or iii) a negative minimum followed by a positive maximum
and then blows up to $-\infty$. Since we are interested in the flows
where the potential is positive along all the way, we have to discard
the third possibility, which occurs when $2\alpha^2>{D\over D-1}$.
As for the first two options, which differ by analoguous inequalities
for $\beta$, they are both acceptable as long as we consider a flow
going from the maximum to the runaway direction at $\phi \to -\infty$.

Although it is possible to solve exactly these flows (for instance choosing
a gauge $b\propto \phi$), the salient features can be extracted
from the asymptotic regimes.
At early times, such a flow will display an inflationary behavior
with de Sitter expansion. The maximum being simple, the flow near it
will be similar to the analytic flows discussed at the beginning of the
section.
At late times, the first term in \potexp\ will dominate and the 
evolution will be quintessence-like like in \quibe, but now
for $t\to \infty$. Depending on the exponent 
$\alpha$, the spacetime will have a de Sitter-like causal structure
in the future if $\gamma>1$ (that is $w<-1+{2\over D}$), or a Minkowski-like
causal structure in the future if $\gamma\leq 1$, as is discussed
in \quint.
The problem however arises of whether it is possible to define
a dual theory. Indeed if we fall short of a future dS phase, we
are unable to relate the dual theory to the future conformal boundary.
{}For the cases with $\gamma\leq 1$, it seems hopeless to try
to think in these terms. For the $\gamma>1$ case, the presence
of a future spacelike boundary is encouraging, but it is clear
that we cannot use there the same (conformal) tools as for a dS boundary.
It would be interesting to understand what kind of UV theory is dual 
to a quintessence cosmology, if a dual theory can be defined at all.
Note that in \townq\ a correspondence for such cases was proposed 
(irrespective of the value of $\gamma$) along the lines of \boonstra. 
The correspondence involves going 
to a ``dual'' frame by making an appropriate rescaling of the
metric with a function of the scalar. The metric is then of dS form, 
however the choice of such a dual frame is rather ad hoc in this set up.

The interest of exponential potentials is that they are likely
to arise in the low energy effective actions derived
from some non supersymmetric versions of string theory.
We turn to this analysis in the following section.

\newsec{Potentials from non supersymmetric strings}
The $D+1$ dimensional low energy effective action of a string theory 
reads in the string frame:
\eqn\leeastr{S = m_s^{D-1}\int d^{D+1}x \sqrt{-\hat{g}}e^{-2\varphi}(
\hat{R}+ 4 \hat{\d}\varphi^2-\VV(\varphi) ).}
We have restricted the fields present to be only the metric and the
($D+1$ dimensional) dilaton.
The action above is derived, for $\VV$ constant, from the requirement
of quantum conformal invariance of the string world-sheet theory.
In a non supersymmetric set up we expect however to have non
vanishing dilaton tadpoles at loop orders. Thus a generic potential
for the dilaton is:
\eqn\dilpot{\VV(\varphi)=\sum_n c_n e^{n\varphi}=c_0 + c_1 e^\varphi +
c_2 e^{2\varphi}+\dots\ . }
The $c_0$ term is related to a tree level cosmological constant,
which arises if the string is in non critical dimensions. Note that
even if we fix the dimension $D$, the sign of $c_0$ is not fixed.
Indeed, allowing for compact directions, the general formula for the
superstring is (see for instance \tsva):
\eqn\noncr{c_0 = {2\over 3} (D+1 +{2\over 3} c_{int} -10),}
where $c_{int}$ is the WS central charge of the ``compact directions''.
Thus starting with an original non critical superstring theory in dimension
greater than 10 and then compactifying down to, say, 4 dimensions, leaves
us with a potential which is positive as $\varphi \to -\infty$.
This has been used in a related context in \silv.

The $c_1$ term is present if the theory is unoriented and/or has an open
string sector, and it has a NSNS tadpole at the disk/crosscap order.
The $c_2$ term originates from a non vanishing torus partition function, 
that is a one loop cosmological constant, which is likely to be present
for a non spacetime supersymmetric theory. Moreover, as noted in \silv, 
a possible further contribution to this order is provided by turning
on RR fluxes on the compact space, so that there is even a little room
for tuning the parameters.

The above considerations clearly need spacetime supersymmetry to be broken
at the string level, however this is perfectly consistent when studying
de Sitter backgrounds, which are incompatible with supersymmetry.
We only need to worry that there are no tachyons in the theory, at least
at tree level. This can be achieved, as for example in the construction
of \silv.

In order to compare with the general discussion of the previous section, 
we go to the Einstein frame by the Weyl rescaling of the metric:
\eqn\weyl{\hat{g}_{\mu\nu}=e^{{4\over D-1}\varphi}{g}_{\mu\nu}.}
The action becomes:
\eqn\actrans{S=m_p^{D-1}\int d^{D+1}x \sqrt{-g}\left(R -{4\over D-1}
\d \varphi^2-e^{{4\over D-1}\varphi}\VV(\varphi)\right),}
where we neglect boundary terms. In order to put the action in the
form \action, we rescale the dilaton by:
\eqn\rescdil{\varphi=\half\sqrt{D-1\over 2}\phi, }
and the potential is:
\eqn\finpot{V(\phi)\equiv e^{{4\over D-1}\varphi}\VV(\varphi) =
\sum_n c_n e^{\left({4\over D-1}+n\right)\half\sqrt{D-1\over 2}\phi}.}

Before analyzing the conditions for such a potential to be
generated by a prepotential, we can already ask which terms can be the
leading term. Indeed, a quintessence-like behaviour is possible provided
the coefficient in the exponential is not too large. It turns out that
the tree level term can always produce quintessence, while the tadpole term
(if it is leading) does so only for $D< 9$. This can be seen as
follows. It is possible to recast a positive potential $V\sim e^{2\alpha
\phi}$ into a prepotential form with $W\sim e^{\alpha\phi}$ if:
\eqn\bound{2\alpha^2< {D\over D-1}.}
The term $n=0$ of \finpot\ always satisfies the bound (which is defined
only for $D>1$). Moreover, from \quibe\ we see that it always leeds
to $\gamma =1$, that is the limiting case for which the expansion
does not produce a horizon.
{}For the term with $n=1$, the bound \bound\ is satisfied only for $D<9$
and the associated $\gamma$ is smaller than 1, that is the expansion
is softer. For instance for $D=3$ it corresponds to an equation
of state with $w=\half$. 
Actually $w_{n=1}$ is always positive. It can never be relevant in the future, 
where it is dominated at least by $w=0$ matter.
This means that if the dilaton potential \finpot\ is to be relevant
to the late stages of the evolution, it has to include a tree level term.

Let us thus write a prepotential that should generate the first three
terms of \finpot, related to those explicitly written in \dilpot.
Take a prepotential of the form \potexp:
\eqn\dilpre{W=Ae^{{1\over \sqrt{2(D-1)}}\phi}-B
e^{{D+1\over 2}{1\over \sqrt{2(D-1)}}\phi}.}
The potential is then computed through \prepot:
\eqn\vdil{V=A^2 e^{\sqrt{2\over D-1}\phi} -AB e^{{D+3\over 2\sqrt{
2(D-1)}}\phi}- B^2 {D-1\over 4} e^{ {D+1\over \sqrt{2(D-1)}}\phi}.}
These can be seen to be the first three terms of \finpot.
The condition for a positive $V$ to derive from a prepotential with a maximum
is thus that:
\eqn\conds{c_0>0, \qquad c_1<0, \qquad c_2<0 \qquad {\rm and}
\qquad c_1^2={D-1\over 4} c_0 |c_2|.}
Although the inequalities above can be satisfied in a suitable
model, it is difficult to think of a symmetry of the WS theory
that would enforce the equality between the constants.\foot{One could
still have a little hope of finding a $c_2$ 
satisfying the equality in \conds\ by tuning the RR fluxes as in \silv, when
this is possible.}
Of course releasing the constraint one still has a potential
with a maximum and a runaway behaviour, but it becomes unaccessible
to our tools of analysis (the solution $W$ to \prepot\ is not
analytic).

Slightly more general potentials were considered in \silv, where
the aim was actually to find not only a positive maximum but also
a positive (or zero) minimum of the potential. This is possible
for $c_0>0$, $c_1<0$ and $c_2>0$, this time satisfying a further
condition resembling to a loosened version of the equality in \conds.
Such a potential could allow in principle flows similar to the one
described at the beginning of the previous section, however it
is not possible to describe them analytically.

\newsec{Discussion}
We have tried in this paper to discuss issues which arise when one
takes seriously the proposal of \stromrg\ that the
cosmological evolution be interpreted as RG flow in a dual Euclidean
field theory associated with the future boundary.
We found particularly useful to borrow from the tools of the AdS/CFT
correspondence the possibility to write first order equations governing
the flow, provided the potential derives from a prepotential. 
In such a preliminary analysis, our perspective has been to review
different kinds of cosmological flows and their associated potentials.
In particular, quantitative investigations on the dual field theory are
still lacking (see however \larsen\ for a recent attempt in this direction).

Having in mind to embed these considerations in string theory,
the necessary lack of supersymmetry\foot{Note that supersymmetric dS vacua
appear in theories with ghosts \refs{\hull,\hullb}. 
{}For these theories the analysis
is much more similar to the AdS case.} 
 prevents us from singling out
specific models in order to extract predictions.
In the context of supergravities, non supersymmetric dS vacua appear
in the potential  
when the theories are ``compactified'' on non-compact manifolds
\refs{\bala,\hullb,\huto,\townq}.
One here can make predictions, however the relevance of these
theories is not yet clear.
More generally, if the supersymmetry is broken to $N=1$ in 4 dimension, 
it is easy to generate potentials with (local) dS minima (see for instance
\kallosh). We are then back to the problem of having too much choice.

If we thus break supersymmetry at the string scale, for instance by
considering a theory in non critical dimension, possibly with 
orientifolds (in order to have a negative tadpole on the crosscap),
we are left with the potentials discussed in the last section.
Here we note that if the dilaton indeed takes the runaway direction, 
and in the presence of a tree level contribution,
the leading behaviour at late times will be $e^a \sim t$, that is
the limiting case for which there is no horizon. If we were to consider
this as a RG flow, we would have the non-trivial problem of even
having to justify the correspondence, since there is no such thing as
a future boundary. Any other term in the dilaton potential
would lead to an even softer behaviour, as noted in \bandin.
On the other hand we could consider models like the one of \silv, 
and flow from the maximum to the minimum. Besides that it cannot be
studied analytically, we have to be aware of the potential problems
in such non supersymmetric theories away from the perturbative
limit $\phi \to -\infty$. Not to mention the degree of fine tuning
(or luck) needed to find a minimum realizing the enormous hierarchy
between the observed cosmological constant and the Planck scale.

As a last remark, one could ask how are to be interpreted flows
from a positive maximum of the potential to a negative minimum.
This situation is puzzling because it clearly admits both dS and 
AdS vacua. One could thus ask whether there can be
a flow mixing both dS and AdS features.\foot{This question 
was raised in conversations with G.~Ferretti and D.~Martelli.}
Considering only time evolution, such a flow will start with
an inflationary phase and then reach the point where the potential
becomes negative. At this point however the scalar cannot possibly
settle at the minimum, since this would imply $H^2<0$ (we assume
flat spacelike sections). Moreover
also an oscillatory behaviour cannot be combined with $\dot{H}<0$.
Since when $V<0$ the scalar field starts to 
violate the dominant energy condition, $p>\rho$, the expansion
is slower and slower and should eventually stop when $H$ reaches zero.
After that the universe would start contracting.
The AdS cosmology with a big crunch is possibly not reached, but the
recontracting behaviour also prevents an holographic interpretation.

\bigskip
\noindent{\bf Acknowledgements:}
I would like to thank G.~Ferretti and L.~Houart for discussions
and interesting comments.
This work is supported by the European Union RTN contract HPRN-CT-2000-00122.

\listrefs

\end